# NeutrSHINE: a high repetition rate ultrafast neutron source driven by SHINE electron beam

Tianyu Ma[1,2,*], Yuchen Liu[1,2,*], Zhangfeng Gao[3], Zuokang Lin[2,†], Hao Li[1,2], Zijian Zhang[2,4],
Zhiyuan Lin[2], Guanchao Wu[5], Yu Zhang[2,4], Yinan Zhu[2], Zhiwen Xu[1,2],
Xinying Jin[6], Weishi Wan[7,‡], and Haixiao Deng[3,§]

[1]*University of Chinese Academy of Sciences, Beijing 100049, China;*
[2]*Shanghai Institute of Applied Physics, Chinese Academy of Sciences, Shanghai 201800, China;*
[3]*Shanghai Advanced Research Institute, Chinese Academy of Sciences, Shanghai, 200120, China;*
[4]*ShanghaiTech University, Shanghai, 201210, China;*
[5]*Institute of Modern Physics, Chinese Academy of Sciences, Lanzhou, 730000, China;*
[6]*Lanzhou University, Lanzhou, 730000, China;*
[7]*Quantum Science Center of Guangdong-HongKong-Macao Greater Bay Area, Shenzhen Guangdong, 518045, China;*

**Abstract:** Neutrons serve as unique "probes" for exploring the microscopic structure of matter, with the performance of a neutron source fundamentally governing the depth of scientific exploration and the breadth of industrial applicability. To address application demands including nuclear data measurement in the ultra-high-energy region, fundamental particle physics research, highly efficient non-destructive neutron testing, and extreme environment simulation, an ultrafast neutron source driven by the 8 GeV electron beam from the Shanghai high-repetition-rate extreme light facility (SHINE) was conceptually proposed, named NeutrSHINE. Using multidisciplinary simulation tools, key neutronic parameters, thermal behavior of high-power neutron targets, and the factors affecting the time resolution of the source were analyzed. The results affirm the technical feasibility and promising application prospects of the NeutrSHINE concept.

## I. INTRODUCTION

Since the 1930s, neutron source technologies have evolved from radioactive sources to reactor-based sources, and further to spallation and laser-driven neutron sources, consistently supporting cutting-edge research in physics, chemistry, materials science, biology, and environmental science. Owing to their neutral charge, strong penetration capability, magnetic moment, sensitivity to light elements, and ability to distinguish isotopes, neutrons are ideal probes for investigating microstructures of matter and are widely used across multiple disciplines. For instance, in fast neutron resonance techniques, neutron characteristics combined with time-of-flight (TOF) methods enable elemental identification through resonance absorption peaks, requiring neutron pulse widths below 100 ps to achieve non-destructive testing precision [1-5]. Neutron source performance depends on beam quality: neutron flux, energy spectrum, and temporal resolution. Current neutron sources face several challenges: spallation sources (such as CSNS Phase-I) can extend the energy spectrum to several hundred MeV and enhance flux, yet their temporal resolution is typically at the hundreds nanosecond level with limited repetition rate [6]; laser-driven neutron sources offer high flux and short pulse duration, but their pulse width remains in the nanosecond range, falling short of the tens of picoseconds resolution required in some cutting-edge applications [7-8]. These limitations hinder progress in fields such as atomic dynamics, neutron capture processes, and radiation effect studies. Hence, developing a neutron source that combines high flux, broad energy coverage, and excellent temporal resolution has become a priority.

---

[*]These authors contributed equally to this work.
[†]Contact author: linzuokang@sinap.ac.cn
[‡]Contact author: wanweishi@quantumsc.cn
[§]Contact author: denghx@sari.ac.cn

Compared to proton-driven sources, electron-accelerator-based neutron sources have historically been constrained by lower flux, a limitation inherent to their underlying physics. Recent advances in high-repetition-rate X-ray free-electron lasers (XFELs), such as the European XFEL [9], LCLS-II [10], and the Shanghai High-repetition-rate X-ray Free-electron Laser Facility (SHINE), have dramatically expanded what is technically achievable. These facilities now support electron accelerators with average beam powers reaching 1 MW and bunch lengths as short as 100 fs. Such capabilities render feasible the development of ultrafast neutron sources driven by high-power electron beams. In particular, SHINE—currently under construction—is centered on an 8 GeV superconducting linear accelerator designed to provide high-quality electron beams. Capitalizing on the design and beam parameters of SHINE, we propose a conceptual ultrafast neutron source, termed NeutrSHINE. This paper presents a theoretical framework to guide its practical realization and to stimulate further research in this emerging area. The proposed design utilizes an 8 GeV electron beam from SHINE to strike a high-Z target (such as tantalum). Through bremsstrahlung and (γ, n) reactions, neutrons are generated via an electron–photon–neutron cascade process, producing a broad energy spectrum ranging from KeV to GeV [11]. Since the bremsstrahlung and (γ, n) processes are much faster than the electron pulse duration, the neutron production timeline is dominated by the electron pulse profile. The theoretical neutron pulse width is approximately 100 fs; although transport processes broaden it to tens of picoseconds, it remains superior to conventional neutron sources. The MHz repetition rate and tens of picosecond pulse duration significantly increase the time-averaged flux while reducing thermal load. NeutrSHINE offers high flux, wide energy spectrum, and ultrafast temporal resolution, opening new pathways for fundamental studies and extreme condition simulations.

The proposed NeutrSHINE offers distinctive advantages in the following areas:

<1> High-Energy Nuclear Data Measurement: Current nuclear data libraries (such as ENDF, JENDL, TENDL) suffer from limited accuracy or missing isotopic data in certain energy ranges due to experimental constraints. For instance, cross-section data for reactions such as $^{232}$Th(n,f) and $^{235}$U(n,f) are severely lacking in the energy region above 100 MeV, which hinders research on ADS transmutation efficiency and the synthesis of superheavy elements. By combining the NeutrSHINE with the time-of-flight (TOF) method, high-precision cross-section measurements in the high-energy region can be achieved, filling critical gaps in nuclear data [12-20].

<2> Research in particle physics: Ultrafast neutrons with energies above 500 MeV can be utilized to investigate transient processes in particle inelastic scattering and to search for rare reaction channels. Additionally, the inelastic scattering at this energy can produce various mesons, serving as an incident beam for meson production [21].

<3> Efficient Neutron Non-Destructive Testing: Conventional neutron non-destructive testing uses neutron-induced gamma rays for elemental imaging, but its broad pulse widths result in high background noise. The NeutrSHINE, with its pulse width of only hundreds of picoseconds, allows for a higher signal-to-noise ratio and more effective testing applications, such as detection of shielded special nuclear materials, explosives, and ammunition assessment [22-23].

<4> Extreme Environment Simulation: Aerospace electronic devices operating in atmospheric radiation environments are susceptible to single-event effects (such as SEU, SEL) caused by secondary ions and recoil nuclei from neutron-induced nuclear reactions. Ground-based simulation of these effects requires a broad-energy neutron spectrum (KeV to GeV) to mimic real atmospheric conditions [24-25].

The NeutrSHINE can generate neutron fluxes far exceeding natural levels, accelerating testing.

This paper is organized as follows: Section II presents the option consideration of NeutrSHINE, including SHINE parameters and layout selection; Section III describes the conceptual design containing main layout and target design; Section IV analyzes neutron parameters covering energy spectrum, angular distribution and time structure; Section V evaluates radiation shielding and target irradiation damage; finally, Section VI summarizes main conclusions and future perspectives [26-30].

## II. OPTION CONSIDERATION OF NEUTRSHINE

SHINE is China's first hard X-ray free-electron laser facility. Construction began in 2018, with the entire system located 29 meters underground and spanning 3.11 kilometers in length. Its 8 GeV superconducting linear accelerator adopts a four-stage acceleration and three-stage compression scheme, achieving a repetition rate of 1 MHz and compressing the pulse width to tens of femtoseconds. It supports simultaneous operation of multiple beamlines [31-36]. The standard parameters of the electron beam it provides are as follows:

TABLE I. Baseline parameters of SHINE electron beam

| Parameters | Injector | LINAC |
|---|---|---|
| Beam energy | 100 MeV | 8 GeV |
| Repetition rate | 1 MHz | 1 MHz |
| Bunch charge | 100-300 pC | 100 pC |
| Pulse length | 1-10 ps | 100 fs |
| Normalized emittance | 0.45 mm-mrad | 0.45 mm-mrad |
| Peak current | 10 A | 1500 A |

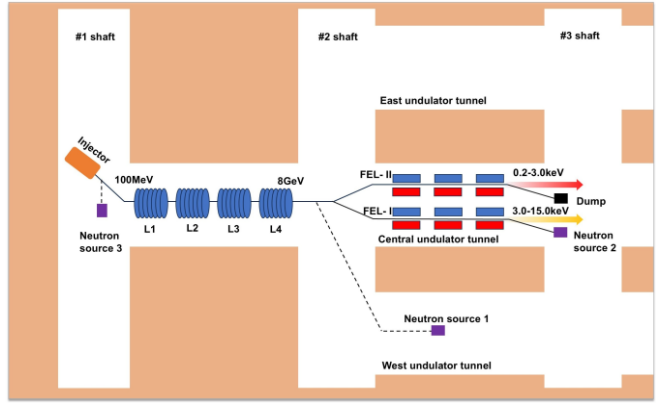

FIG 1. The layout of SHINE accelerator complex and tunnels.

FIG 1 illustrates the beam distribution section layout. SHINE's linear accelerator, with a total length of 1.4 km, accelerates the electron beam from 100 MeV to 8 GeV through four-stage acceleration and three-stage compression, reducing the pulse width to tens of femtoseconds. The 8 GeV electron beam at 1 MHz repetition rate is then directed into a distribution system for multi-beamline operation. The NeutrSHINE layout follows an "electron generation → photon conversion → neutron production → beam transport → application stations" logic. Design must consider constraints from existing tunnel and shaft structures. Preliminary options include:

<1> Constructing a hundred-MeV neutron source using electron beams extracted at the injector end. The electron beam with 100 MeV energy, 0.1 mA average current and Gaussian beam size of 18 mm incident on a 5×5×6 cm$^3$ tantalum target. Monte Carlo simulation show that the 6 cm thick target absorbs 93.47 MeV of energy, with a total neutron escape rate of 1.98×10$^{13}$ n/s. Neutron beams under these parameters can be used for neutron radiography, activation analysis, etc. The neutron energy spectrum is shown in FIG 2.

<2> Guiding the electron beam to the East (West) undulator tunnel (currently unused) for target bombardment. This offers greater spatial flexibility and facilitates overall neutron source layout, though subsequent experimental areas need expansion on the same level as the accelerator.

<3> Utilizing spent beams from FEL-I and FEL-II in the beam dump at the bottom of Shaft 3. This option maintains beam quality and has minimal impact on neutron production. However, it requires downward beam guidance and expansion of underground space.

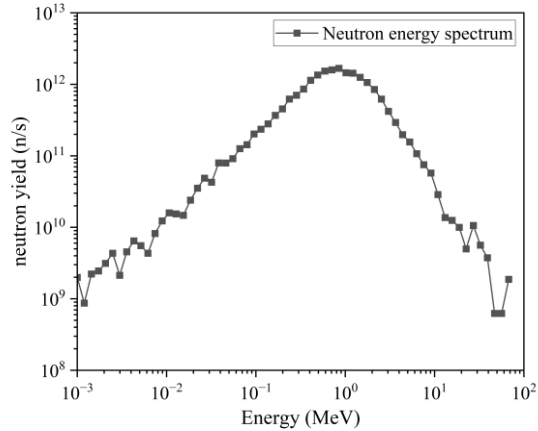

FIG 2. Neutron energy spectrum curve of 100 MeV electron incident.

The following sections focus on the conceptual design of the 8 GeV beam driven NeutrSHINE.

## III. CONCEPTUAL DESIGN OF NEUTRSHINE

**A. Main Layout**

The overall layout adopts a regular octagonal structure, comprising a photoneutron conversion target station, neutron beamlines, shielding components, and experimental endstations (FIG 3):

Electron Beam Injection Zone: The 8 GeV electron beam is deflected by bending magnets onto the high-Z target. A beam expansion device controls the spot size to reduce power density and ensure target safety.

Photon-neutron conversion zone: Electrons interact with the high-Z target, generating bremsstrahlung photons and neutrons via cascade reactions. Neutrons are nearly isotropic and are collimated into different beamlines. The target station employs a disk-style water cooling structure to maintain temperatures below the boiling point of water. Moderators can be added to adjust the neutron energy spectrum from keV to GeV [37-38]. The initial yield peaks in the 1 MeV region, with single-pulse flux reaching up to $10^{16}$ cm$^{-2}$·s$^{-1}$.

Hot cell for target exchanging: The target assembly is a slender block with integrated cooling channels and shielding. A hot cell equipped with robotic arms, water cooling circuits, and spare target assemblies enables automated replacement (FIG 3.).

Beam transmission and application zone: The inner side of the octagon accommodates one electron beamline and seven neutron beamlines, each equipped with superconducting magnets, time-of-flight spectrometers, and sample environment systems (FIG 4.).

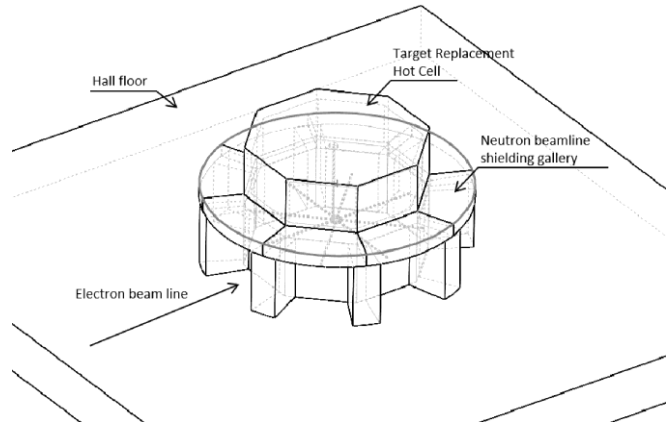

FIG 3. Schematic diagram of the target station.

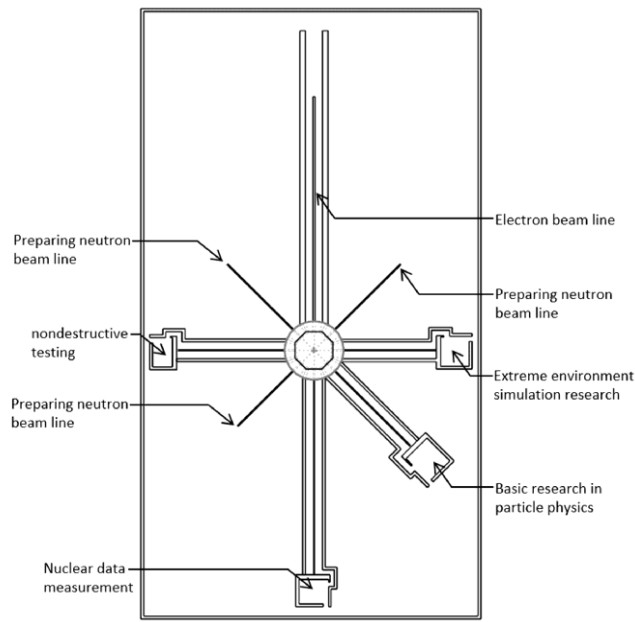

FIG 4. Horizontal cross-section of the target station.

For example, in nuclear data measurements: the endstation is located 65 meters from the NeutrSHINE. The beamline includes a size-reducing collimator I, an electron-removing magnet (to reduce background), and an adjustable-aperture collimator II (with movable copper pins featuring 1–5 cm holes) to accommodate different experimental requirements. The distances and positions of other terminals will be designed as required.

**B. Neutron Target Design**

Current electron beam size provided by SHINE is 1×1 mm$^2$, which can be scanned into square beam spots with different size. In thermal and yield analysis, it can be equivalent to a uniform beam of the same size. Therefore, a right square prism tantalum target was designed, leveraging the material's high atomic number, corrosion resistance, and machinability. Its neutronics parameters were simulated with MCNPX$^{TM}$, enabling coupled physical and thermal analysis through integration with ANSYS® FLUENT. For a right square prism target (defined by base length and thickness) under perpendicular beam incidence, the determination of them is described below.

The MCNPX results ($R_{psp}$) are normalized per source particle. The actual average value $R_{avg}$ is obtained by multiplying $R_{psp}$ by the average current 0.1 mA ($6.2422\times10^{14}$ e/s). Thus, the instantaneous value $R_{ins}$ is given by scaling $R_{psp}$ by the peak current 1000 A (100 pC/100 fs). Unless otherwise specified, all subsequent results refer to the average value $R_{avg}$.

An 8 GeV, 11×11 cm² uniform electron beam incident on the bottom base of the target was simulated by MCNPX™. The target was modeled with a fixed base of 13×13 cm² and a variable thickness, and the coordinate origin was set at the center of its bottom base, with the beam incident along the positive Z-axis. The average neutron flux within a 1 cm radius was measured by point detectors at (50, 0, 3), (30, 0, 40), and (0, 0, 50), as well as the total neutron yield.

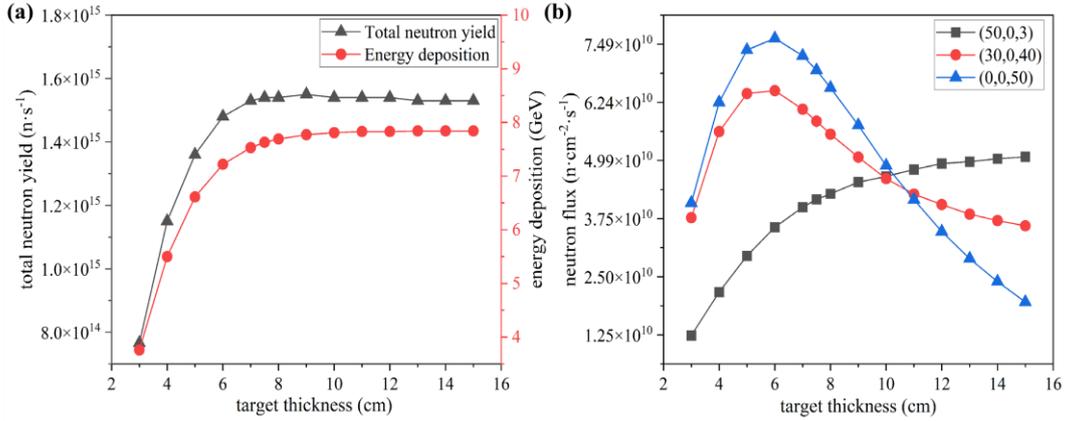

FIG 5. (a) Total neutron yield and energy deposition vs. target thickness. (b) Neutron flux vs. target thickness at various detector locations.

FIG 5. (a) shows that the energy deposition and total neutron yield saturate beyond a target thickness of 7 cm. FIG 5. (b) shows that increasing the thickness from 6 cm to 7 cm enhanced the flux at (50, 0, 3) by 12.06%, despite a 5.28% reduction at (0, 0, 50). Since a greater thickness would negatively impact the time resolution of neutrons from the top surface, and considering the results in FIG 5 (a), a thickness of 7 cm was therefore selected as the optimal compromise.

With the target thickness fixed at 7 cm, various target base dimensions and beam spot sizes were analyzed. A 1 cm lateral extension of the target base effectively reduces doses leakage. The energy deposition and neutron escape data, obtained from MCNPX™ simulations, and the peak temperature (T) in the core region, as computed by FLUENT based on the cooling structure shown in FIG 6. (a), are summarized in TABLE II for each beam spot size. Simulation results indicate that varying the side length of the target base has a negligible impact on neutron yield, but it significantly affects the angular distribution of neutrons in space.

TABLE II. Effect of beam spot size on selected physical parameters.

| Base size (cm²) | 12×12 | 13×13 | 14×14 | 15×15 | 16×16 | 17×17 |
|---|---|---|---|---|---|---|
| Beam size (cm²) | 10×10 | 11×11 | 12×12 | 13×13 | 14×14 | 15×15 |
| Total escape (n/e) | 2.475 | 2.472 | 2.456 | 2.461 | 2.438 | 2.437 |
| Lateral escape (n/e) | 1.284 | 1.214 | 1.141 | 1.090 | 1.018 | 0.978 |
| D (W/m³) | $2.26\times10^9$ | $1.94\times10^9$ | $1.59\times10^9$ | $1.39\times10^9$ | $1.22\times10^9$ | $1.09\times10^9$ |
| T (°C) | 118 | 104 | 91.1 | 81.8 | 74.7 | 69.2 |

Effective neutron target cooling requires dissipating heat from the high-energy deposition core while avoiding coolant boiling for safe operation. The corresponding cooling structure is presented in FIG 6. (a).

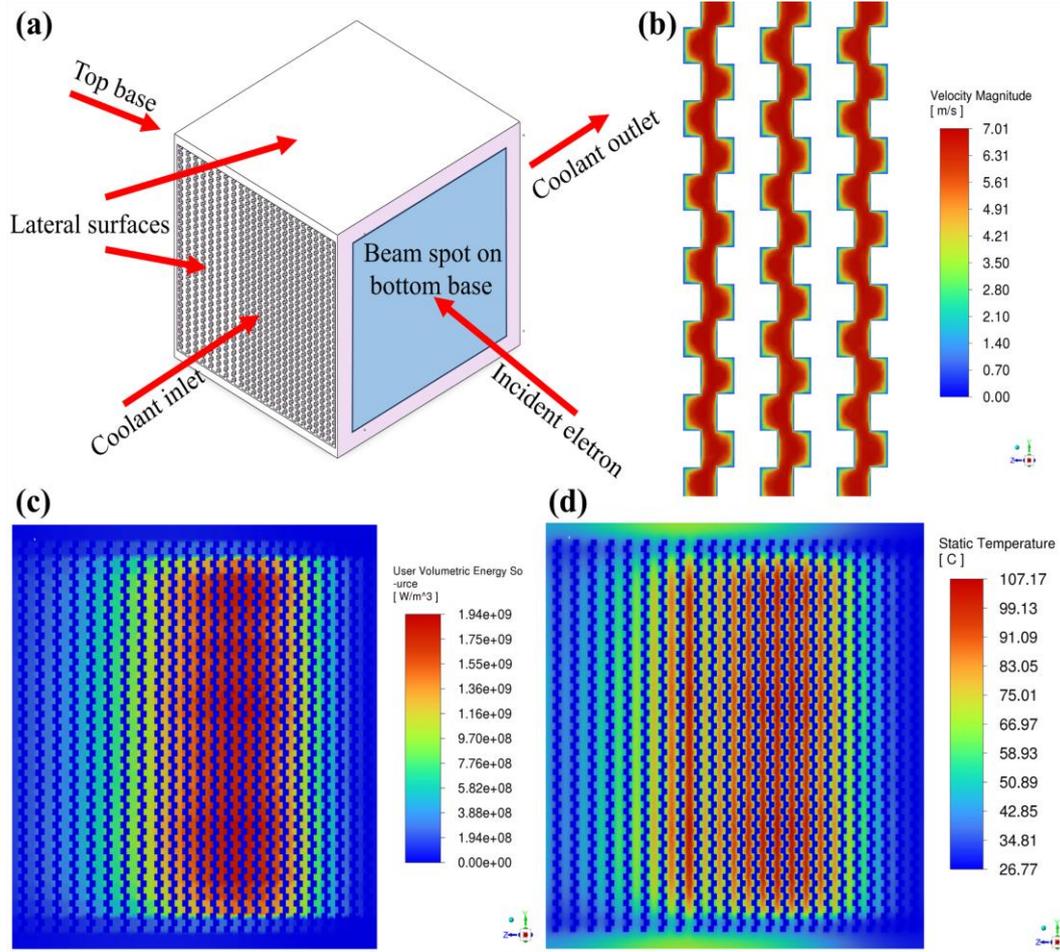

FIG 6. (a) Target cooling structure. (b) Coolant velocity distribution on YOZ plane. (c) Energy deposition density on YOZ plane. (d) Temperature distribution on YOZ plane.

The disk-style cooling structure in the target's core (FIG 6. (a)) consists of a stack of 25 tantalum sheets and 24 water layers. The sheet thicknesses vary sequentially as 2.5 mm (1st), 3.5 mm (2nd–16th), 4.5 mm (17th–24th), and 3 mm (25th). Each sheet contains 1×2 mm² grooves (2 mm pitch) that are aligned across water channels, yielding total equivalent thicknesses of 70 mm (Ta) and 48 mm (water). This design increases heat transfer area and breaks up the hydrodynamic boundary layer. A uniform flow field with straight inlets/outlets was assumed for simulation simplicity, and the pressure difference between the inlet and outlet is 0.3 MPa.

The final design of a 13×13×7 cm³ target for an 11×11 cm² beam spot was determined based on thermal and safety analyses. Simulations confirm effective heat dissipation (FIG 6. (d)), with a maximum core temperature of 107.2°C and a sub-cooled interface at 96°C, which can be further secured by pressurizing the system.

## IV. NEUTRON PARAMETERS OF NEUTRSHINE

### A. Energy spectrum and angular distribution

TABLE III provides a brief summary of the physical parameters of SHINE Ultrafast Neutron Source. FIG 7. shows the energy spectrum from NeutrSHINE exhibits a continuous distribution from keV to GeV, with several distinct characteristic photon peaks visible. And FIG 8. and FIG 9. (a) shows neutron escape is slightly more pronounced in the forward direction (Z-axis) than laterally. And FIG 9. (b) shows the forward-peaked angular distribution of the photons.

TABLE III. Physical parameters of the NeutrSHINE.

| Target material | Tantalum |
|---|---|
| Base size | 13×13 cm$^2$ |
| Target thickness (non-aquifer) | 7 cm |
| Target thickness (aquifer) | 11.8 cm |
| Spot size | 11×11 cm$^2$ |
| Peak neutron yield | 1.53×10$^{22}$ n/s |
| Average neutron yield | 1.53×10$^{15}$ n/s |
| Highest energy deposition density | 1.94×10$^9$ W/m$^3$ |

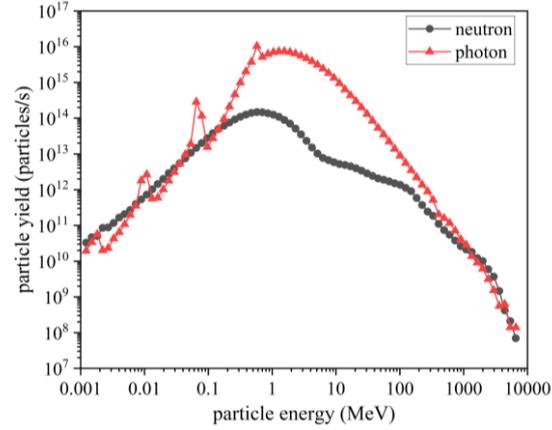

FIG 7. Energy spectrum of neutron and photon.

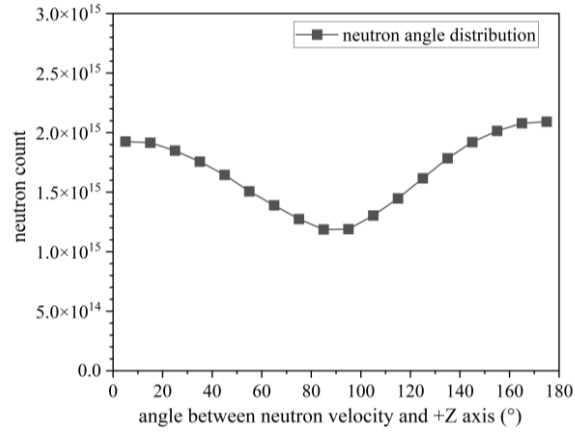

FIG 8. Neutron angle distribution.

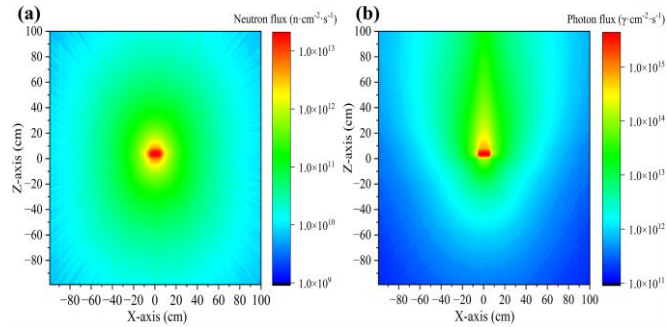

FIG 9. (a) Distribution of neutron flux in XOZ cross-section. (b) Distribution of photon flux in XOZ cross-section.

TABLE IV compares key parameters of the CSNS-I, the East China Accelerator Neutron Source (ECANS) [39], and the NeutrSHINE.

TABLE IV. Parameter comparison of selected neutron sources.

| Neutron source | CSNS-I | ECANS | NeutrSHINE |
|---|---|---|---|
| Types of target particles | Proton | Electron | Electron |
| Energy of target particles | 1.6GeV | 35MeV | 8GeV |
| Average beam current | 62.5μA | 2mA | 0.1mA |
| Beam pulse length | ＜400ns |  | 100fs |
| Target material | tungsten | tungsten | tantalum |
| Repetition frequency | 25Hz | 100Hz | 1MHz |
| Target cooling capacity | 100kW | 100kW | 800kW |
| Neutron energy | 1eV-200MeV | 1eV-30MeV | 1keV-8GeV |
| Neutron yield | $10^{16}$/s | $10^{10}$-$10^{13}$/s | $10^{15}$/s |
| Neutron pulse length | ns | ns | ~100ps |
| Neutron ratios of different energies | 53％(1eV-1MeV)<br>40％(1MeV-20MeV)<br>5％(20MeV-200MeV) | 0.01％(＜1eV)<br>1.20％(1eV-1keV)<br>21.44％(1keV-100keV)<br>47.79％(100keV-1MeV)<br>27.63％(1MeV-5MeV)<br>1.88％(5MeV-30MeV) | 5.61％(1keV-100keV)<br>65.19％(100keV-1MeV)<br>27.01％(1MeV-10MeV)<br>1.98％(10MeV-100MeV)<br>0.21％(100MeV-8GeV) |

**B. Time structure**

The neutron beam pulse length is determined by three factors: the incident electron pulse duration, the neutron generation depth, and the intrinsic moderation effect. Since the latter two effects are difficult to isolate, their combined influence is attributed to the target size [40-41].

Temporal analysis was performed on neutrons from the top base for two reasons: partial cancellation of electron and neutron transit times through the target, and the significant pulse broadening for lateral neutrons caused by a large (11×11 cm²) beam spot.

For the SHINE electron beam (Gaussian profile, 0-100 fs, 65 fs FWHM), the standard deviation of the arrival time distribution for neutrons reaching the top base across various energy ranges is summarized in Table V.

Table V. The temporal structure of neutrons arriving at top base in different energy ranges.

| Neutron energy | Average arrival time (ns) | Standard deviation (ns) |
|---|---|---|
| 0.01MeV | 48.10 | 35.40 |
| 0.1MeV | 21.06 | 14.35 |
| 0.5MeV | 7.699 | 4.412 |
| 1MeV | 4.047 | 2.306 |
| 5MeV | 1.703 | 0.764 |
| 10MeV | 1.298 | 0.5148 |
| 50MeV | 0.7455 | 0.1915 |
| 100MeV | 0.6010 | 0.1232 |
| 500MeV | 0.3845 | 0.0355 |
| 1000MeV | 0.3483 | 0.0233 |

Table V lists a ±10% energy range (e.g., 100 MeV represents the 90-110 MeV bin). Results show that higher neutron energies yield shorter pulses. For 500 MeV neutrons, the standard deviation is ~35.5 ps, corresponding to a FWHM of 83.6 ps (assuming a Gaussian shape). This value is significantly lower than the temporal resolution of other international neutron sources.

The pulse shapes for different neutron energies are shown in FIG 10, which indicate a certain degree of overlap in the arrival times of neutrons across different energies. On a positive note, this overlap is expected to diminish as the distance between the target and the application end-station increases.

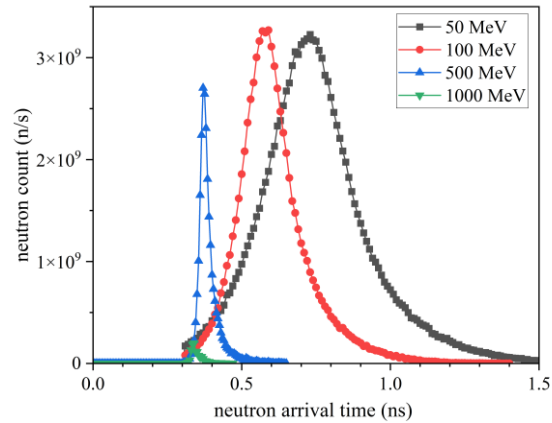

FIG 10. Distribution of arrival times of neutrons of different energies.

## V. RADIATION SHIELDING AND IRRADIATION DAMAGE OF THE TARGET

This section quantitatively assesses the radiation safety and service life of the target station's core components via MCNPX$^{TM}$ simulations, focusing on external dose levels outside the shielding and irradiation damage to the target itself.

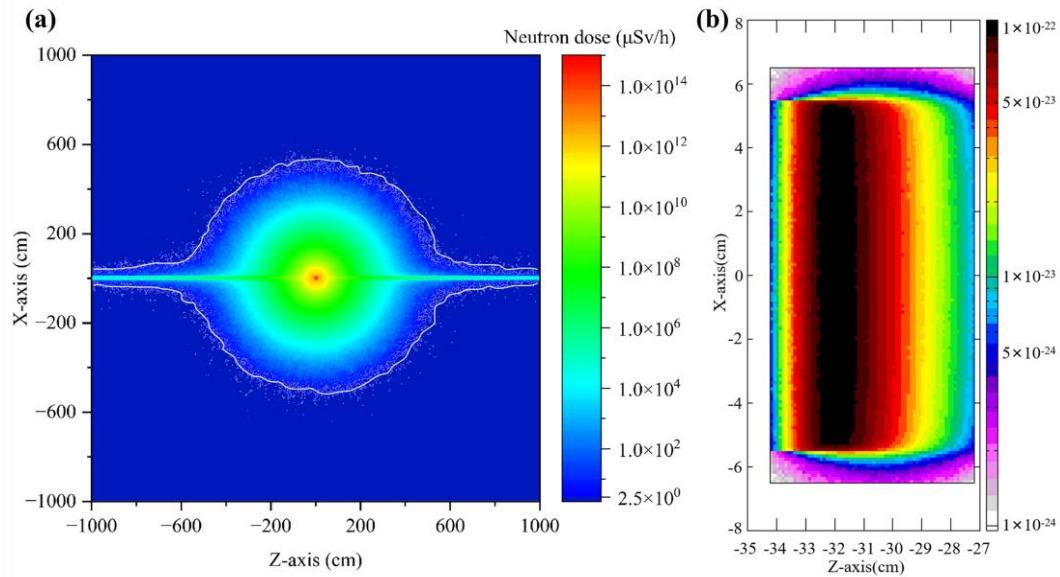

FIG 11. (a) Instantaneous neutron radiation dose distribution. (b) Tantalum target radiation damage.

Simulations of a regular octagonal prism shielding structure composed of 6 m of Q235B steel (FIG 9 (a)) confirm its effectiveness in neutron radiation protection. FIG 11 (a) shows the neutron dose distribution around the operational station, with the white contour marking the 2.5 μSv/h limit, well under the 20 mSv/year safety standard set by the International Atomic Energy Agency (IAEA) [42-43].

The radiation damage rate in the tantalum target was simulated with FLUKA to assess operational longevity. FIG 11 (b) shows the maximum displacement per atom (DPA) value of the target was $1.0817\times10^{-22}$ DPA/s. Based on the DPA limit of 11 for tantalum, the service life t of the neutron target is calculated as:

$$t = \frac{11}{1.0817\times10^{-22} \times 6.2422\times10^{14} \times 365\times 24\times 3600} \approx 5.17(\text{year}) \quad (1)$$

In summary, the analysis confirms that the shielding design contains radiation within safe limits and maintains acceptable component damage, ensuring the target station's long-term operational safety and stability.

## VI. CONCLUSION

In this study, we present the design of an ultrafast neutron source based on the 8 GeV electron beam from the SHINE facility, incorporating a 13×13×7 cm³ tantalum target. Its feasibility is confirmed through multidisciplinary simulations. Thermal-hydraulic analysis performed with FLUENT indicates that a coolant flow velocity of approximately 7 m/s maintains the target core temperature below 107.2 °C and the interface below 96 °C, thereby preventing water boiling and confirming the effectiveness of the disc-cooling structure. The design produces a high-flux neutron beam with a broad energy spectrum and delivers neutron pulses that are substantially shorter than those available at most existing international facilities. As an example, the FWHM pulse duration for 500 MeV neutrons reaches 83.6 ps, providing strong support for research on ultrafast dynamics. Radiation shielding design keeps the dose below 2.5 μSv/h, in full compliance with IAEA safety standards. It is important to note that the electron beam quality remains adequate for driving the neutron source following XFEL lasing. Under optimized conditions, both XFEL and neutron sources can therefore be operated concurrently to serve user experiments. In summary, the proposed neutron source demonstrates strong performance in terms of user requirement, neutron output, thermal management, and radiological safety, laying a solid foundation for the engineering implementation of a high-repetition-rate neutron source at SHINE and opening avenues for future experimental platforms.

**Acknowledgment**

This work was supported by the National Natural Science Foundation of China (12125508), the National Key Research and Development Program of China (2024YFA1612104), the CAS Project for Young Scientists in Basic Research (YSBR-042), the Research and Development of Key Technologies for High-Frequency and High-voltage Proton Accelerators – Neutron Target Research Project (E455110102)，and Shanghai Pilot Program for Basic Research – Chinese Academy of Sciences, Shanghai Branch (JCYJ-SHFY-2021-010).

**References**

[1] W. Zhou, et al. Development of the double spoke cavity prototype for CSNS-II. Nucl. Instrum. Methods Phys. Res., Sect. A. **1062**, 169170(2024).

[2] H.-W. Wang, et al. Development of photo-neutron facility driven by electron LINAC. Nuclear Techniques. **37**, 10 (2014).

[3] N. Kardjilov, I. Manke, R. Woracek, A. Hilger, and J. Banhart. Advances in neutron imaging. Mater. Today, **21**, 6 (2018).

[4] Z.-Y. Lin, Z.-K. Lin, Z.-J. Zhang, G.-C. Wu, Y.-N. Zhu, Y.-C. Liu, Y. Dai, and Z.-M. Dai, Analysis of the influence of boron distribution on compound biological effectiveness values in boron neutron

capture therapy. Int. J. Radiat. Oncol., Biol., Phys. **122**, 2(2025).

[5] Z.-J. Zhang, Z.-K. Lin, G.-C. Wu, Z.-Y. Lin, Y.-C. Liu, T.-Y. Ma, H. Li, Y. Dai, and Z.-M Dai, Design and analysis of an AB-BNCT self-shielding rotating neutron target. Nucl. Eng. Technol. **57**, 9(2025).

[6] J. Bao, et al. Experimental result of back-streaming white neutron beam characterization at Chinese spallation neutron source. Acta Phys. Sin. **68**, 8(2019).

[7] I. Pomerantz, et al. Ultrashort pulsed neutron source. Phys. Rev. Lett. **113**, 184801(2014).

[8] I. Pomerantz, et al. Laser generation of ultra-short neutron bursts from high atomic number converters. In *Proceedings of Laser Acceleration of Electrons, Protons, and Ions III; and Medical* Applications of Laser-Generated Beams of Particles III (Vol. 9514, pp. 54-58). SPIE.

[9] T. Tschentscher, C. Bressler, J. Grünert, A. Madsen, A. P. Mancuso, M. Meyer, A. Scherz, H. Sinn, and U. Zastrau. Photon beam transport and scientific instruments at the European XFEL. Appl. Sci. **7**, 592(2017).

[10] J. N. Galayda, et al. (LCLS-II Collaboration), The new LCLS-II project: Status and challenges. In Proceedings of LINAC2014, TUIOA04, 404(2014).

[11] X.-J. Jiao, et al. A tabletop, ultrashort pulse photoneutron source driven by electrons from laser wakefield acceleration. Matter Radiat. Extremes, **2**, 6(2017).

[12] D. A. Brown, et al. ENDF/B-VIII.0: the 8th major release of the nuclear reaction data library with CIELO-project cross sections, new standards and thermal scattering data. Nucl. Data Sheets, **148**, 1(2018).

[13] K. Shibata, et al. JENDL-4.0: a new library for nuclear science and engineering. J. Nucl. Sci. Technol. **48**, 1(2011).

[14] A. J. Koning, D. Rochman, J. C. Sublet, N. Dzysiuk, M. Fleming, and S. van der Marck, TENDL: complete nuclear data library for innovative nuclear science and technology. Nucl. Data Sheets, **155**, 1(2019).

[15] Z.-Z. Ren, et al. Measurement of the $^{232}$Th(n, f) cross section in the 1-200 MeV range at the CSNS Back-n. Nucl. Sci. Tech. 34, 115(2023).

[16] A. Manna, et al. (n_TOF Collaboration), New insights on fission of $^{235}$U induced by high energy neutrons from a new measurement at n_TOF. Phys. Lett. B, **860**, 139213(2025).

[17] Y.-J. Qiu, et al. Measurement of the $^{239}$Pu(n, f) cross section from 4 keV to 100 MeV using the white neutron source at the CSNS Back-n facility. Phys. Rev. C, **107**, 024606(2023).

[18] G.-L. Yang, et al. (The CSNS Back-n Collaboration), Measurement of the neutron capture cross sections of rhenium up to stellar s- and r- process temperatures at the China Spallation Neutron Source Back-n facility. Phys. Rev. Res. **6**, 013225(2024).

[19] M. Calviani, et al. (n_TOF Collaboration), High-accuracy $^{233}$U(n, f) cross-section measurement at the white-neutron source n_TOF from near-thermal to 1 MeV neutron energy. Phys. Rev. C, **80**, 044604(2009).

[20] A. C. Larsen, A. Spyrou, S. N. Liddick, and M. Guttormsen, Novel techniques for constraining neutron-capture rates relevant for r-process heavy-element nucleosynthesis. Prog. Part. Nucl. Phys. 107, 69(2019).

[21] D. Akimov, et al. (COHERENT Collaboration), Simulating the neutrino flux from the Spallation Neutron Source for the COHERENT experiment. Phys. Rev. D **106**, 032003(2022).

[22] A. Favalli, et al. Demonstration of active neutron interrogation of special nuclear materials using a high-intensity short-pulse-laser-driven neutron source. Sci. Rep. **15**, 724(2025).

[23] B.-H. Zhao, Y.-T. Zhou, C. Xie, and Y.-Z. Zeng, Research on the application of neutron detection in the identification of the effectiveness of munitions. Journal of Innovation and Development, **10**, 1(2025).

[24] Q.-Z. Yu, Evaluating the broad neutron spectrum of ANIS. Appl. Radiat. Isot. **203**, 111075(2024).

[25] Z.-B. Wu, Y.-L. Yao, X.-F. Shen, K. Li, A.-D. Liu, X.-T. He, and B. Qiao, Study of the atmospheric neutron radiation effects using compact laser-driven spallation neutron source. Phys. Rev. Res. **7**, 013071(2025).

[26] D. B. Pelowitz, et al. MCNPX$^{TM}$ User's Manual, Version 2.7.0. Los Alamos National Laboratory Tech. Rep. LA-CP-11-00438(2024). https://mcnpx.lanl.gov/

[27] T. Sato, et al. Recent improvements of the particle and heavy ion transport code system–PHITS version 3.33. J. Nucl. Sci. Technol. **61**, 1(2024).

[28] Ansys® Academic Research Mechanical, Ansys Student 2024 R1. https://www.ansys.com/

[29] C. Ahdida, et al. New capabilities of the FLUKA multi-purpose code. Front. Phys. **9**, 788253(2022).

[30] G. Battistoni, et al. Overview of the FLUKA code. Ann. Nucl. Energy, **82**, 10(2015).

[31] W. Decking, et al. A MHz-repetition-rate hard X-ray free-electron laser driven by a superconducting linear accelerator. Nat. Photonics **14**, 650 (2020).

[32] F. Zhou, et al. Commissioning of the SLAC Linac Coherent Light Source II electron source, Phys. Rev. Accel. Beams **24**, 073401 (2021).

[33] K. Li, and H.-X. Deng, Systematic design and three-dimensional simulation of X-ray FEL oscillator for Shanghai Coherent Light Facility. Nucl. Instrum. Methods Phys. Res., Sect. A, 895, 40(2018).

[34] N.-S. Huang, H.-X. Deng, B. Liu, D. Wang, and Z.-T. Zhao, Features and futures of X-ray free-electron lasers. The Innovation, **2**, 2(2021).

[35] T. Liu, et al. Status and future of the soft X-ray free-electron laser line at the SHINE, Front. Phys., **11** (2023) 1172368.

[36] N.-S. Huang, et al. The MING proposal at SHINE: megahertz cavity enhanced X-ray generation, Nucl. Sci. Tech. **34**, 6(2023).

[37] Y.-N. Zhu, Z.-K. Lin, H.-Y. Yu, X.-H. Yu, and Z.-M. Dai, Conceptional design of an adjustable moderator for BNCT based on a neutron source of 2.8 MeV proton bombarding with Li target. Nucl. Eng. Technol. **56**, 5(2024).

[38] Y.-N. Zhu, Z.-K. Lin, H.-Y. Yu, and X.-H. Yu, A study on the treatment of brain tumors with BNCT using several moderators with different thicknesses. Appl. Radiat. Isot. **208**, 111303(2024).

[39] Y.-Y. Wu, B.-W. Cai, J.-J. Zou, Y.-W. Yang, J.-H. Li, J.-F. Wan, X.-C. Peng, Y. Liu , D.-X. Xiao, and B. Tang, Simulation analysis of 35 MeV high-power electron accelerator driven white neutron source target. Appl. Radiat. Isot. 217, 111632(2025).

[40] J. Zweiback, T. E. Cowan, R. A. Smith, J. H. Hartley, R. Howell, C. A. Steinke, G. Hays, K. B. Wharton, J. K. Crane, and T. Ditmire, Characterization of fusion burn time in exploding deuterium cluster plasmas. Phys. Rev. Lett. **85**, 3640(2000).

[41] J.-Y. Tang, H.-T. Jing, C. Zhang, Z. Yang, H.-H. Xiao, H.-Q. Tang, Z.-Y. Zhou, X.-C. Ruan Q.-W. Zhang, Key nuclear data measurements for advanced fission energy and white neutron source at CSNS. At. Energy Sci. Technol. 47, 7(2013).

[42] N. Cherkashyna, D. D. DiJulio, T. Panzner, E. Rantsiou, U. Filges, G. Ehlers, P. M. Bentley, Benchmarking shielding simulations for an accelerator-driven spallation neutron source. Phys. Rev. ST Accel. Beams **18**, 083501(2015).

[43] G.-C. Wu, Z.-K. Lin, Z.-J. Zhang, Z.-Y. Lin, Y.-N. Zhu, Y. Dai, and Z.-M. Dai, Neutron activation

dose assessment based on a human head phantom post-BNCT. Health Phys. Doi:10.1097/HP.0000000000001968(2025).